\documentclass[twocolumn]{autart}   
\usepackage{graphicx}         
\usepackage{lineno,hyperref}
\usepackage{amsmath}
\usepackage{setspace}
\usepackage{float}
\usepackage{color}
\usepackage{longtable}
\usepackage{subfig}
\usepackage[table]{xcolor}
\usepackage{nicefrac}
\usepackage{epstopdf}
\usepackage{enumerate}
\usepackage{booktabs, multicol, multirow}
\usepackage{mathtools}
\usepackage{tikz}
\usepackage{url}
\usepackage{hyperref}
\usetikzlibrary{shapes.geometric, arrows,calc}
\newtheorem{remark}{Remark} 
\newtheorem{theorem}{Theorem}

\tikzstyle{startstop} = [rectangle, rounded corners, minimum width=3cm, minimum height=0.6cm,text centered, draw=black]

\tikzstyle{io} = [trapezium, trapezium left angle=70, trapezium right angle=110, minimum width=1cm, minimum height=0.6cm, text centered, draw=black, text width=2.5cm]

\tikzstyle{smalproc} = [rectangle, minimum width=0.2cm, minimum height=0.6cm, text centered,text width=2.5cm, draw=black]

\tikzstyle{process} = [rectangle, minimum width=3cm, minimum height=0.6cm, text centered, text width=9cm, draw=black]

\tikzstyle{decision} = [diamond,aspect=2.5, minimum width=0.1cm, minimum height=0.4cm,text width=2.5cm, text centered, draw=black]

\tikzstyle{textonly} = [aspect=2.5, minimum width=0.1cm, minimum height=0.4cm,text width=2.5cm, text centered]

\tikzstyle{arrow} = [thick,->,>=stealth]

\begin{document}
\begin{frontmatter}

\title{Optimal Power Distribution Control for a Network of Fuel Cell Stacks}

\author[iitm]{Resmi Suresh M P},    
\author[gdpl]{Ganesh Sankaran},               
\author[gdpl]{Sreeram Joopudi},  
\author[iitm]{ Shankar Narasimhan}, 
\author[nmrl]{ Suman Roy Choudhury},
\author[iitm]{Raghunathan Rengaswamy}\ead{raghur@iitm.ac.in}  

\address[iitm] {Indian Institute of Technology, Madras, Chennai, Tamil Nadu, India 600036.}  
\address[gdpl] {Gyan Data Private Limited, Taramani, Chennai, India}
\address[nmrl] {Naval Materials Research Laboratory, Ambernath, India}

\begin{abstract}                          
In power networks where multiple fuel cell stacks are employed to deliver the required power, optimal sharing of the power demand between different stacks is an important problem. This is because the total current collectively produced by all the stacks is directly proportional to the fuel utilization, through stoichiometry. As a result, one would like to produce the required power while minimizing the total current produced. In this paper, an optimization formulation is proposed for this power distribution control problem. An algorithm that identifies the globally optimal solution for this problem is developed.  Through an analysis of the KKT conditions, the solution to the optimization problem is decomposed into off-line and on-line computations. The on-line computations reduce to simple equation solving. For an application with a specific v-i function derived from data, we show that analytical solutions exist for on-line  computations. We also discuss the wider applicability of the proposed approach for similar problems in other domains.
\end{abstract}
\end{frontmatter}
\section{Introduction}
Fuel cells have gained considerable attention in the field of power conversion, especially in the past few decades \cite{rohland1992hydrogen,carrette2001fuel,spiegel2011pem}. Fuel cells are capable of extracting energy from the fuel more efficiently than internal combustion engines \cite{carrette2001fuel}. Due to this higher fuel-efficiency, less harmful emissions, low operational noise and absence of moving parts, fuel cells have been proposed for a wide variety of applications \cite{larminie2003fuel,carrette2001fuel}. The power obtainable from a single fuel cell is rather low in comparison to the power requirement for most of the applications. Hence, to meet realistic power requirements, active surface area has to be increased. In view of this, multiple fuel cells are stacked or connected in series and/or parallel arrangements to enhance the output voltage and power \cite{handbook2004eg,howstuffworks}. 

\begin{figure}[htbp]
\centering
\includegraphics[scale=0.25, trim=0.8in 0.8in 0.5in 0.5in, clip=true]{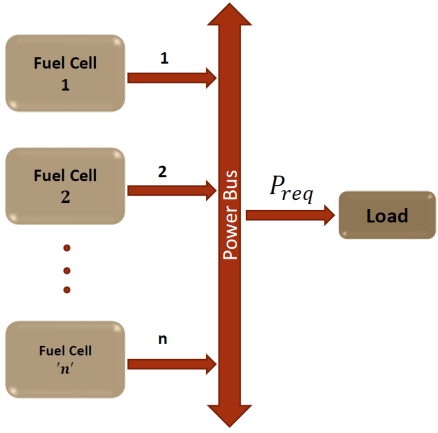}
\caption{Schematic of a parallel connected fuel cell network with n branches}
\label{fuelcellconfig}
\end{figure}

While fuel cell technology has several advantages, there are also problems related to long term durability and gradual loss of performance over time. The performance degradation could be due to many factors such as electrolyte degradation, catalyst poisoning, water flooding etc. \cite{carrette2001fuel,handbook2004eg}. Such degradation can result in reduced current being drawn from the stack at the same voltage. As a result, the overall system performance could be compromised, and in the worst case of complete stack failure, the whole system could fail \cite{handbook2004eg}. This issue is usually addressed by designing a reliable parallel connected fuel cell architecture with an additional power bus as shown in Fig. \ref{fuelcellconfig} \cite{michalskeparallel:10}. Another approach to improve reliability is the use of multiple power sources in the same network \cite{jiang2003strategy,jiang2007adaptive}. While considerable research has been devoted to concerns related to output voltage, state of charge, reliability and network configuration, very little work has focused on developing operational strategies that minimize the fuel consumption through online control. We address this control problem in this paper.  
\section{Problem Statement} 
\begin{figure}[htbp]
\centering
\includegraphics[scale=0.25, trim=1in 0in 0.35in 0.45in, clip=true]{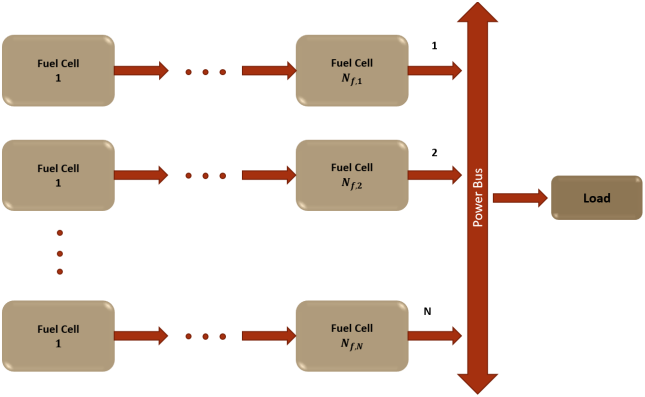}
\caption{Schematic of a parallel connected fuel cell network with N branches and $N_{f,i}$ fuel cell stacks in the $i^{th}$ branch.}
\label{fuelcellconfig_all}
\end{figure}

Consider a fuel cell network with N branches and $N_{f,i}$ stacks in the $i^{th}$ branch (as shown in Fig. \ref{fuelcellconfig_all}). The control problem is one of minimizing the total current drawn (a surrogate for total fuel utilized) while meeting a fixed power requirement. This can be posed as the following optimization problem.
\begin{subequations}
\label{genrl_optimeqn}
\begin{align}
& \underset{I_{i}}{\text{min}}
& & I_{net} = \sum_{i=1}^{N}I_{i} \\
& \text{subject to} & &  \sum_{i=1}^{N}\sum_{j=1}^{N_{f,i}} P_{ij} = P_{req} \\
& & &  V_{ij} \geq V_{ij,lb} \quad \forall \ i=1:N; \ j=1:N_{f,i} & \\
& & &  V_{ij} \leq V_{ij,ub} \quad\forall \ i=1:N; \ j=1:N_{f,i} &   
\end{align}
\end{subequations} 
where $I_{net}$ is the total current from all branches, $I_{i}$ represents the current drawn from the $i^{th}$ branch of the fuel cell network. $V_{ij}$, $P_{ij}$ $\left(= \phi_{ij}V_{ij}I_{ij}\right)$ and $\phi_{ij}$ are voltage, power produced and efficiency of the $j^{th}$ stack in the $i^{th}$ branch respectively. Current ($I_i$) drawn from each branch is the decision variable. Subscripts $lb$ and $ub$ denote the lower and upper bounds respectively. 
\vskip-0.1in
\textit{Assumptions}:

\begin{itemize}
\item[1] The potential across each fuel cell is assumed to be a static map $f_{ij}$: 
\begin{equation}
V_{ij}= f_{ij}(I_i)\qquad \quad
\forall \quad i=1:N; \quad j=1:N_{f,i}
\label{staticmap}
\end{equation}
\item[2] Number of stacks in each branch of the network under consideration, is assumed to be one ($N_{f,i} = 1 \quad \forall \ i=1:N$).
\item[3] Minimum possible power from the fuel cell network is achieved when all stacks are operating at their corresponding $I_{lb}$. Pathological cases where $P_{ij}|_{I_{i,ub}} < P_{ij}|_{I_{i,lb}}$ for $j^{th}$ stack in $i^{th}$, which may arise depending on the voltage-current profile, are neglected. 
\end{itemize}
Using equation \eqref{staticmap}, voltage constraints can be transformed to current constraints. Also, a network with multiple stacks in a branch can be converted to an equivalent network with a single stack in each branch, as current flow in each stack of a branch is equal. Power in each branch of the resulting equivalent network can be written as 
\begin{equation}
P_i = P_{eff,i}  = \phi_i V_i I_i   =  \phi_i f_i(I_i) I_i 
\label{pow}
\end{equation}
where $V_i$, $P_i$ and $\phi_i$ are the voltage, power deliverable and efficiency of the single fuel cell stack in the $i^{th}$ branch of the equivalent circuit.

Using all these assumptions, the general constrained optimization problem described in (P1) can be reduced to the following minimization problem (P2) for a network as shown in Fig. \ref{fuelcellconfig}.
\begin{subequations}
\label{optim_final}
\begin{align}
& \underset{I_{i}}{\text{min}}
& & I_{net} = \sum_{i=1}^{N}I_{i}  \\
& \text{subject to} & &  \sum_{i=1}^{N} P_{i} = \sum_{i=1}^{N} \phi_i f_i(I_i) I_{i} = P_{req} \label{eq_const}\\
& & & I_{i,lb}-I_{i} \leq 0 \qquad \quad \forall \quad i=1:N & &\\
& & & I_{i} -I_{i,ub}\leq 0 \qquad \quad \forall \quad i=1:N & &
\end{align}
\end{subequations}
The optimization problem (P2) in equation \eqref{optim_final} involves a linear objective function with one nonlinear equality constraint and 2N linear inequality constraints. Due to non-linear equality constraint present,this optimization problem becomes non-convex. 

\section{Theoretical Analysis using KKT Conditions}
\label{genrlsoln}

\begin{theorem}
\label{theorem1}
When a fuel cell network is used to achieve a power of $P_{req}$, optimum is such that stacks operating at a current value within allowable limits ($I_{i,lb} < I_i < I_{i,ub}$) are maintained at equal $\frac{dP_{i}}{dI_i}$. 
\end{theorem}

\textbf{Proof:} 
For any V-I characteristics, Karush-Kuhn-Tucker (KKT) conditions \cite{kuhn1951proceedings} can be used to prove the above theorem. Lagrangian of the minimization problem described in \eqref{optim_final} can be written as follows.
\begin{multline}
L = \sum_{i=1}^N I_i +\lambda \left(P_{req} -\sum_{i=1}^N P_i\right) + \sum_{i=1}^{N} \mu_i (I_{i,lb} - I_i)\\ + \sum_{i=1}^{N}\gamma_i (I_i - I_{i,ub})
\end{multline}
There are 2N inequality equations out of which first N are lower bounds on $I_i$ and the last N are upper bounds on $I_i$. Necessary conditions for optimality as described by KKT conditions for all $i \ \epsilon \ \{1,...,N\}$ are
\begin{gather}
\mu_i \geq 0 \\
\gamma_i \geq 0\\
\label{kktcondn1}
\mu_i (I_{i,lb}-I_i) = 0  \\ \label{kktcondn2}
\gamma_i (I_i - I_{i,ub}) = 0
\end{gather}
Different cases in KKT conditions are obtained by varying the number of active and inactive constrains, all of which falls into any of the following 2 cases for different values of M and K. 
\subsection*{Case 1}
For $1 \leq M \leq N$ and $1 \leq K \leq N$, \\
Constraints $1:M$ are inactive, $M+1:N$ are active.\\
Constraints $N+1:N+K$ are inactive, $N+K+1:2N$ are active.
\begin{enumerate}[(a)]
\item If $K \leq M$, for $i = M+1:N$, $I_i = I_{i,lb} = I_{i,ub}$ which is not feasible.
\item If $K > M$, for $i = K+1:N$, $I_i = I_{i,lb} = I_{i,ub}$ which is not feasible.
\end{enumerate}
\subsection*{Case 2}
For $1 \leq M \leq N$ and $1 \leq K \leq N$, \\
Constraints $1:M$ are inactive, $M+1:N$ are active.\\
Constraints $N+1:N+K$ are active, $N+K+1:2N$ are inactive.
\begin{enumerate}[(a)]
\item If $K>M$, 
For $i = 1+M:K$, $I_i = I_{i,lb} = I_{i,ub}$ which is not feasible.

\item If $K\leq M$,
\begin{equation*}
I_i = \left\{ \,
\begin{tabular}{c c}
$I_{i,ub} $ & for $i=1:K$ \\
$I_{i,lb}$ & for $i=1+M:N$ 
\end{tabular}
\right.
\end{equation*}
Using $\frac{dL}{dI_i} = 0$, 
\begin{equation}
\label{dldi}
1 - \lambda \left(\frac{dP_i}{dI_i}\right) - \mu_i +\gamma_i = 0 \qquad \forall \ i=1:N
\end{equation}
For the given case, KKT conditions can be reduced to
\begin{gather*}
\mu_i = \left\{ \,
\begin{tabular}{c c}
0 & for $i=1:M$ \\
$\geq 0$ & for $i=1+M:N$ 
\end{tabular}
\right.\\
\gamma_i = \left\{ \,
\begin{tabular}{c c}
$\geq 0$ & for $i=1:K$ \\
0 & for $i=1+K:N$ 
\end{tabular}
\right.
\end{gather*}
Substituting in equation \eqref{dldi}, 
\begin{gather}
\label{condn1}
1 - \lambda \left( \frac{dP_i}{dI_i} \right)_{I_{i,ub}} +\  \gamma_i = 0 \qquad \forall \quad i = 1:K \\ \label{condn2}
1 - \lambda \left( \frac{dP_i}{dI_i} \right)_{I_{i,opt}} = 0 \qquad \forall \quad i = 1+K:M \\
\label{condn3}
1 - \lambda \left( \frac{dP_i}{dI_i} \right)_{I_{i,lb}}- \mu_i = 0 \ \ \forall \ \ i = 1+M:N 
\end{gather}
$I_{i,opt}$ is the optimum current drawn from $i^{th}$ stack.From equation \eqref{condn2},
\begin{equation}
\lambda =  \frac{1}{\left( \frac{dP_i}{dI_i} \right)_{I_{i,opt}}} \qquad \forall \quad i = K+1:M 
\end{equation}
\begin{multline}
\implies
\left( \frac{dP_{k+1}}{dI_{k+1}}\right)_{I_{k+1,opt}}=\left( \frac{dP_{k+2}}{dI_{k+2}}\right)_{I_{k+2,opt}}=... \\ =\left( \frac{dP_{M}}{dI_{M}}\right)_{I_{M,opt}}  \label{equaldpdi_all}
\end{multline}
Hence, $\frac{dP_i}{dI_i}$ of all stacks operating within their allowable limits are equal. Also, from  \eqref{condn1} and \eqref{condn3},
\begin{multline}
\gamma_j = -1 + \frac{\left( \frac{dP_j}{dI_j} \right)_{I_{j,ub}}}{\left( \frac{dP_i}{dI_i} \right)_{I_{i,opt}}} \quad \forall \quad j=1:K; \\ i = K+1:M 
\end{multline}
\begin{multline}
\mu_j = 1 - \frac{\left( \frac{dP_j}{dI_j} \right)_{I_{j,lb}}}{\left( \frac{dP_i}{dI_i} \right)_{I_{i,opt}}} \qquad \forall \quad j=1+M:N; \\ i = K+1:M 
\end{multline}

It can be seen that $\gamma_j \geq 0$ is satisfied if and only if
\begin{equation}
\label{condn4}
\left( \frac{dP_j}{dI_j} \right)_{I_{j,ub}} \geq \left( \frac{dP_i}{dI_i} \right)_{I_{i,opt}}
\end{equation}
and $\mu_j \geq 0$ can be achieved only if
\begin{equation}
\label{condn5}
\left( \frac{dP_j}{dI_j} \right)_{I_{j,lb}} \leq \left( \frac{dP_i}{dI_i} \right)_{I_{i,opt}}
\end{equation}
Combining \eqref{condn4} and \eqref{condn5}, 
\begin{gather}
\label{finalcondn}
\left( \frac{dP_j}{dI_j} \right)_{I_{j,lb}} \leq \left( \frac{dP_i}{dI_i} \right)_{I_{i,opt}} \leq \left( \frac{dP_r}{dI_r} \right)_{I_{r,ub}} \\
j = M+1:N; \ \ i = K+1:M; \quad r = 1:K
\end{gather}
\end{enumerate}
Thus, first order optimality conditions imply that the only feasible solution to the concerned optimization problem (P2) is such that 'M' stacks are operating at a current higher than their minimum allowable current ($I_{i,lb}$) and out of which 'K' stacks are operating at $I_{i,ub}$ and the conditions \eqref{equaldpdi_all} and \eqref{finalcondn} are satisfied. 

\section{Proposed Algorithm}
\label{algorithm}
A new algorithm is proposed to find an optimum solution to the minimization problem (P2) described using equation \eqref{optim_final}. 
For a network of 'N' branches with a single stack in each branch, to deliver $P_{req}$, say $ I_i > I_{i,lb}$ for first 'm' stacks and among these, 'k' stacks operate at $ I_i = I_{i,ub}$ at optimum. Stacks $k+1$ to $m$ are then operating at a current in between their bounds, $I_{i,lb} < I_i < I_{i,ub}$. At optimum, according to theorem \ref{theorem1},
\begin{equation}
\label{equaldpdi}
\left(\frac{dP_{i}}{dI_i}\right)_{I_{i,opt}} = \left(\frac{dP_{j}}{dI_j}\right)_{I_{j,opt}}  \quad \forall \quad i,j =k+1:m\\
\end{equation}
subject to
\begin{subequations}
\label{allcondnskkt}
\begin{gather}
\label{preq}
\sum_{i=k+1}^{m}P_{i}=P_{req}^{eff}\\
\label{inequaldpdi}
 \left(\frac{dP_{j}}{dI_j}\right)_{_{I_{j,lb}}} \leq \left(\frac{dP_{i}}{dI_i}\right)_{_{I_{i,opt}}} \leq \left(\frac{dP_{r}}{dI_r}\right)_{_{I_{r,ub}}} 
\end{gather}
\end{subequations}
\begin{equation*}
\label{dpdimax}
\text{such that} \quad j=m+1:N;\quad i=k+1:m;\quad  r=1:k
\end{equation*}
and
\begin{equation}
\label{effectv_power}
P_{req}^{eff} = P_{req} - \sum_{j=m+1}^{N} P_{j,min} - \sum_{r=1}^k P_{r,max} 
\end{equation}
Corresponding to a power requirement of $P_{req}$, let the optimal current drawn from branches be $I_1$, $I_2$,..., \text{and} $I_N$, then according to the algorithm, any incremental power demand $dP_{net}$ should be met by drawing additional current from those '$l$' such that,
\begin{equation}
\left(\frac{dP_{net}}{dI_l}\right)_{I_l} \geq \left(\frac{dP_{net}}{dI_n}\right)_{I_n} \quad \forall \ \ n \ \epsilon \ N
\end{equation}
For a power requirement of $P_{req}$, any configuration that satisfies the above criterion will be a local optimum to the problem. Multiple solutions are possible depending on the V-I profile. Global optimum can be obtained based on the value of objective function for those solutions which satisfy all these constraints. The solution which has lowest value of objective function will be the global optimum. This optimum will have maximum rise in total power from the network for a specific increase in current drawn $\left(\left( \frac{dP_{net}}{dI_{net}} \right ) = \sum_{i=1}^{N}  \frac{dP_{i}}{dI_i}\right)$. 
\begin{remark}
The optimization problem described in \eqref{optim_final} is convex for any V-I profile, $V_i = f_i(I_i)$, with a restriction that corresponding power ($P_i = \phi_i \ I_i \ f_i(I_i)$) is a concave function (or in other words Hessian of power is negative definite). For this convex subproblem, though multiple solutions are obtained by solving equations \eqref{equaldpdi}, \eqref{allcondnskkt} and \eqref{effectv_power}, only one of them will satisfy the actual constraints of the optimization problem (P2), resulting in a unique solution. Proof can be found in Appendix \ref{red.optim.pbm}.
\end{remark}

\section{Implementation}
\begin{figure}[h]
\begin{tikzpicture}[scale=0.7, every node/.style={scale=0.65},node distance=1.65cm,    line/.style={
      draw,thick,
      -latex',
      shorten >=2pt
    }]
\node (start) [startstop, yshift=-1.65cm] {Start};
\node (in1) [io, below of=start, yshift=0.5cm] { $f(I_i)$, $I_{ub}$, $I_{lb}$};
\node (pro1) [process, below of=in1, yshift=0.35cm] {Calculate $P_{obt,min}$ and $P_{obt,max}$};
\node (pro2a) [process, below of=pro1, yshift=-0cm] {Arrange cells in decreasing order of $(\frac{dP}{dI})$ calculated for each cell at $I_{i,lb}$ and $I_{i,ub}$. $i=1:2N$ represents index of sorted list};
\node (pro3) [process, below of=pro2a, yshift=-0cm] {Compute $P_{obt}$ at each observable points};
\node (dec1) [decision, below of=pro3, yshift=-0.3cm] { $P_{obt,min} \leq P_{req} \leq P_{obt,max}$};
\node (inp2) [io, left of=dec1, yshift=0cm, xshift = - 3cm] {$P_{req}$ };\node (pro2b) [io, right of=dec1, xshift=3.2cm] {Error: Required power cannot be obtained};
\node (pro4) [smalproc, below of=dec1, yshift=-0.5cm] {n=n+1 };
\node (dec2) [decision, below of=pro4, yshift=-0.1cm] {$P_{req} \leq P_{obt,n}$};
\node (pro5a) [process, below of=dec2, yshift=-0.3cm] {Find cells operating at $I_{ub}$ (k cells) and $I_{lb}$ ($N-m$ cells) };
\node (pro5b) [smalproc, right of=dec2, xshift=3.2cm] {Update n};
\node (pro5c) [process, below of=pro5a, yshift=-0.1cm] {Find cell numbers for which current needs to be optimized ($m-k$ cells)};
\node (pro6) [smalproc, below of=pro5c, yshift=0.1cm] {Calculate $P_{req}^{eff}$ \eqref{effectv_power}};
\node (pro7) [process, below of=pro6, yshift=-0cm] {Solve\ $(\frac{dP_{i}}{dI_i})=constant$ \ $\forall \quad i=k+1:m$ \\and \ $\sum_{i=1}^{k}P_i=P_{req}^{eff}$ };
\node (pro8) [process, below of=pro7, yshift=-0.1cm] {Current in cells $i=k+1:m$};
\node (pro9) [process, below of=pro8, yshift=-0.1cm] {All possible local optima};
\node (out1) [io, below of=pro9, yshift=0.2cm] {Optimum Current: $I_{opt}$};
\node (stop) [startstop, below of=out1,yshift = 0.5cm] {Stop};
\node(tex1)[textonly, below of=out1, yshift=0.28cm, xshift = 5cm] {Online mode};
\draw [arrow] (start) -- (in1);
\draw [arrow] (in1) -- (pro1);
\draw [arrow] (pro1) -- (pro2a);
\draw [arrow] (pro2a) -- (pro3);
\draw [arrow] (pro3) -- (dec1);
\draw [arrow] (inp2) -- (dec1);
\draw [arrow] (dec1) -- node[anchor=east] {n=0} node[anchor=west]{yes} (pro4);
\draw [arrow] (dec1) -- node[anchor=south] {no} (pro2b);
\draw [arrow] (pro4) -- (dec2);
\draw [arrow] (dec2) --node[anchor=east] {yes} (pro5a);
\draw [arrow] (dec2) --node[anchor=east] {yes} node[anchor=west] {Corresponding observable point = n-1} (pro5a);
\draw [arrow] (dec2) --node[anchor=south] {no} (pro5b);
\draw [arrow] (pro5b) |- (pro4);
\draw [arrow] (pro5a) -- (pro5c);
\draw [arrow] (pro5c) -- (pro6);
\draw [arrow] (pro6) -- (pro7);
\draw [arrow] (pro7) -- node[anchor=east] {Multiple solutions} (pro8);
\draw [arrow] (pro8) -- node[anchor=east] {$Current\ \forall \ i=1:k \ \& \ i=m+1:N$}(pro9);
\draw [arrow] (pro9) -- node[anchor=east] {Minimum total current}(out1);
\draw [arrow] (out1) -- (stop);
\draw[magenta,thick,dotted]($(inp2.north west)+(-1.4,0.6)$)  rectangle ($(out1.south east)+(5.85,-1.1)$);
\end{tikzpicture}
\caption{Proposed algorithm: A fuel cell network with N branches is considered such that, to deliver the required power ($P_req$) m cells should be operated at a current higher than $I_{lb}$, out of which k cells operate at $I_{ub}$.}\label{algo}
\end{figure}
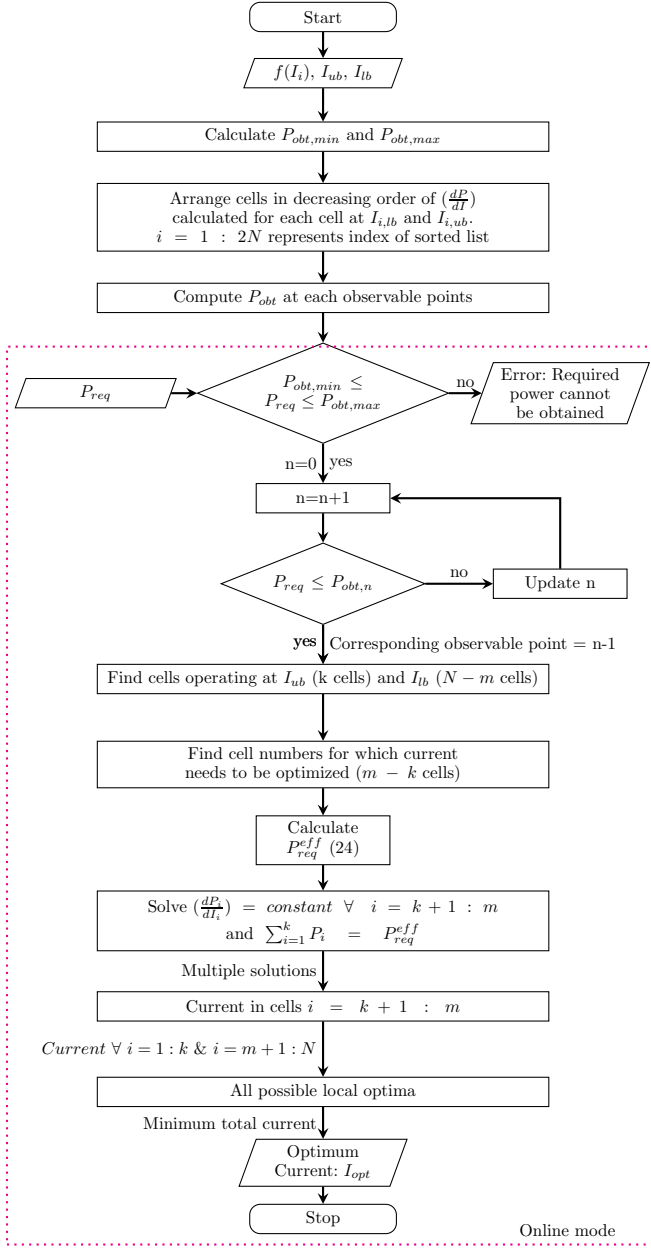
The implementation of proposed algorithm is described in Fig. \ref{algo}. Input to the algorithm will be the V-I model parameters, bounds on current and the required power ($P_{req}$). Using the new algorithm, majority of the calculations for estimating optimum current can be performed off-line, reducing the on-line computations. The major step in the off-line mode is to identify observable points (kink points which decide whether to add or remove stacks from the list of stacks, for which current needs to be optimized). Observable points are decided based on $(\nicefrac{dP_{j}} {dI_j})$ of each stack evaluated at both $I_{j,lb}$ and $I_{j,ub}$, where j corresponds to the branch number. Each value in the list of $(\nicefrac{dP_{j}} {dI_j})$ corresponds to an observable point. For a fuel cell network with N branches, there will be 2N observable points. These observable points are then arranged such that the power corresponding to the observable point decreases down the list. Corresponding to the $i^{th}$ value of $(\nicefrac{dP} {dI})$ in the sorted list, if for any $j \ \epsilon \ 1:N$, $(\nicefrac{dP_{j}}{dI_j})_{_{I_j= I_{j,lb}}} \leq (\nicefrac{dP}{dI})_i$, then $j^{th}$ stack should be operating at $I_{j,lb}$. And for any  $j \ \epsilon \ 1:N$, if $(\nicefrac{dP_{j}}{dI_j})_{_{I_j=I_{j,lb}}} > (\nicefrac{dP}{dI})_i > (\nicefrac{dP_{j}}{dI_j})_{_{I_j=I_{j,ub}}}$, then $j^{th}$ stack should be operating at a current in between $I_{j,lb}$ and $I_{j,ub}$. Optimum current to be drawn from these stacks can be obtained by solving $\left(\nicefrac{dP_j}{dI_j}\right)_{I_{j,opt}} = (\nicefrac{dP}{dI})_i$. Further, if for any $j \ \epsilon \ 1:N$, $(\nicefrac{dP_{j}}{dI_j})_{_{I_j= I_{j,ub}}} \geq (\nicefrac{dP}{dI})_i$, then $j^{th}$ stack should be operating at $I_{j,ub}$. Hence, corresponding to each observable point, current in each stack is pre-calculated off-line. This information can be used to evaluate the power at every observable point.

When there is a specific power requirement, the stacks to be operated at current values between their bounds are identified based on the pre-calculated power at each observable point. Upper and lower bound constraints on current are inactive for these stacks. The optimum current to be drawn from each of these stacks is evaluated on-line such that equations \eqref{equaldpdi}, \eqref{allcondnskkt} and \eqref{effectv_power} are satisfied. The exact solution to this set of equations has been obtained analytically for a specific function $ V_i = f_i(I_i)$ and is discussed in section \ref{anlyt}. If multiple solutions are obtained by solving these set of equations which satisfy the bounds on current, then global optimum is found based on the value of objective function. The solution with lowest value of net current is the global optimum for the problem.
 

\section{Analytical solution for a specific convex  subproblem}
\label{anlyt}
\begin{figure}[htbp]
\centering
\includegraphics[scale=0.75]{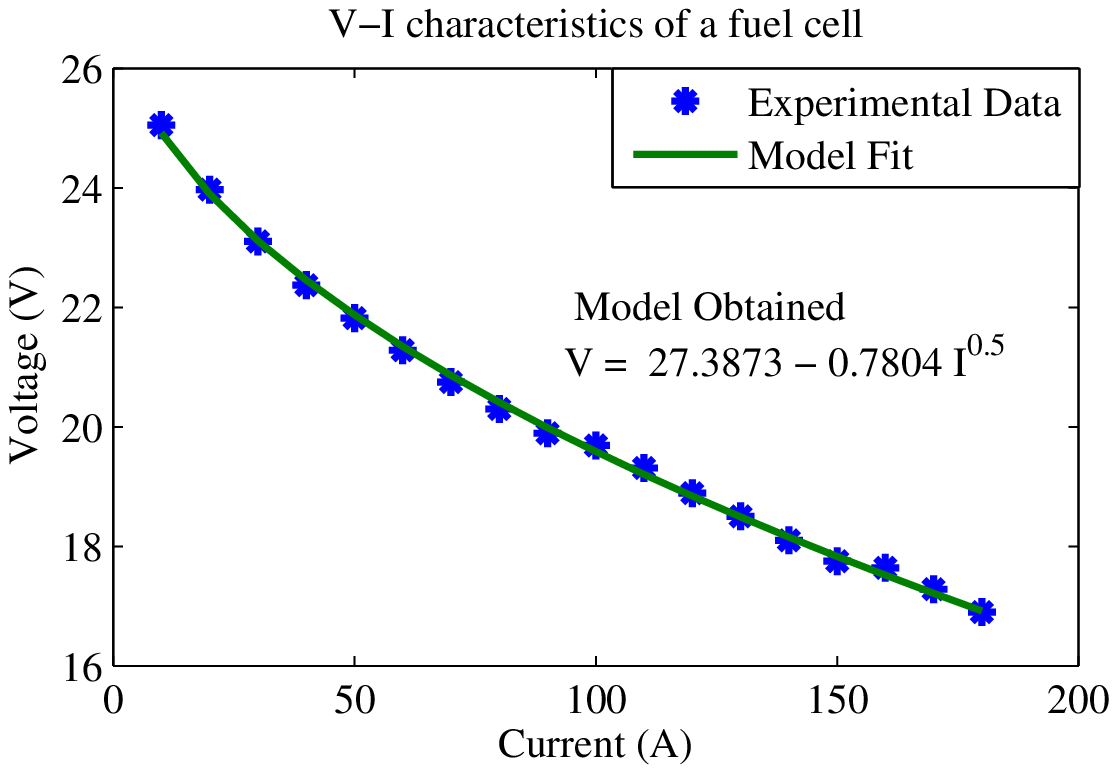}
\caption{Comparison of experimental data obtained from NMRL with the assumed V-I characteristics form.}
\label{examplevi}
\vskip-0.2cm
\end{figure}
Voltage and current data (See Fig. \ref{examplevi}) for a typical fuel cell stack are obtained from Naval Materials Research Laboratory (NMRL). V-I characteristics of the form as described in \eqref{voltage profile} is assumed for which corresponding power is a concave function. The assumed profile has been tested on the data and a $99.87 \%$ good fit is obtained as shown in Fig. \ref{examplevi}.
\begin{subequations}
\label{voltage profile}
\begin{align}
V(I_i) = f_i(I_i) &= a_i + b_i \ \sqrt{I_i} \\
\text{such that} \quad a_{i} &\geq 0 \\
b_{i} &\leq 0 \qquad \forall \quad i=1:N &
\end{align}
\end{subequations}
As power corresponding to this V-I profile is concave, corresponding optimization problem is convex as described in Appendix \ref{red.optim.pbm} and thus will have a unique global solution. Analytical solution is derived for this specific V-I characteristics. 

Consider '$n1$' cells operating at $I_{j,opt}$ at optimum such that $I_{j,lb}<I_{j,opt}<I_{j,ub}$. The following equations can be obtained using Theorem 1 and are used to determine optimum current drawn from those 'n1' cells as shown in Fig. \ref{algo}. 
\begin{equation}
P_1 + P_2 + ... + P_{n1} = P_{req}^{eff}
\label{totP}
\end{equation}
\begin{equation}
\frac{dP_1}{dI_1} = \frac{dP_2}{dI_2} = ... =\frac{dP_{n1}}{dI_{n1}}
\label{dpdiopt}
\end{equation}
Using equation \eqref{pow} and \eqref{voltage profile}, and substituting $x_j = \sqrt{I_j}$, equation \eqref{totP} and \eqref{dpdiopt} reduces to
\begin{equation}
\sum_{j=1}^{n1}\left(\phi_{j} a_j+\phi_j b_{j}x_j \right)x_j^2 = P_{req}^{eff}
\label{totp1}
\end{equation}
\begin{equation}
 \left(\phi_1 a_{1}+\frac{3}{2} \phi_1 b_{1}x_1 \right)=   \left( \phi_j a_{j}+\frac{3}{2} \phi_j b_{j}x_j \right) \quad \forall \quad j = 2:n1
 \label{phixeq}
\end{equation}
For $j = 1:n1$, rearranging equation \eqref{phixeq},
\begin{gather}
x_j = g_j x_1 + h_j  \label{xvalues}\\
x_{j}^2 = g_j^2 x_1^2 + 2 g_j h_j x_1 + h_j^2\\
x_j^3 = g_j^3 x_1^3 + x_1^2 \left(3g_j^2 h_j\right) + x_1 \left(3g_j h_j^2\right) + h_j^3
\end{gather}
where
\begin{equation}
\quad g_j = \frac{\phi_1 b_{1}}{\phi_j b_{j}} \quad and \quad
h_j = \frac{\left( \phi_1 a_{1} - \phi_j a_{j} \right)}{1.5\phi_j b_{j}}
\end{equation}
Substituting in equation \eqref{totp1} and simplifying, we get a cubic equation. 
\begin{multline}
x_1^3 \left(\sum_{j=1}^{n1}\phi_j b_{j}g_j^3 \right) + 
x_1^2 \left(\sum_{j=1}^{n1}\left(3\phi_j b_{j}g_j^2 h_j + \phi_j a_{j}g_j^2 \right) \right)\\ +x_1 \left(\sum_{j=1}^{n1} \left(3\phi_j b_{j}g_j h_j^2 + 2\phi_j a_{j}g_j h_j \right) \right) =P_{req}^{eff} 
\label{cubiceqn}
\end{multline}
Current drawn from all stacks can be evaluated using equations \eqref{xvalues} and \eqref{current_eqn} after solving for $x_1$ (using equation \eqref{cubiceqn}).
\begin{equation}
\label{current_eqn}
I_{opt,j} = x_j^2
\end{equation}
As the analytical solution is obtained by solving a cubic equation (Equation \eqref{cubiceqn}), it will result in three sets of solutions. But as the problem is convex and has a unique minimum, it is guaranteed that there can only be one solution set among those three which will satisfy all constraints of the actual optimization problem (P2). 
\section{Results and Discussion}
The algorithm proposed in section \ref{algorithm} has been implemented on various fuel cell networks. All fuel cell stacks are assumed to follow V-I characteristics described in equation \eqref{voltage profile}. Parameters for various stacks are obtained by perturbing the parameters calculated for a typical fuel cell stack V-I profile shown in Fig. \ref{examplevi}. The results obtained for three different case studies are presented here. The working of algorithm is shown using the first case study while the second one aims at demonstrating the application of algorithm for a network with multiple stacks in a branch. Finally, efficiency of the algorithm is compared with that of standard optimizers. 
\begin{figure*}[htbp]
\centering
\includegraphics[scale=0.38,trim = 0.9cm 1.5cm 0cm 0cm,clip=true]{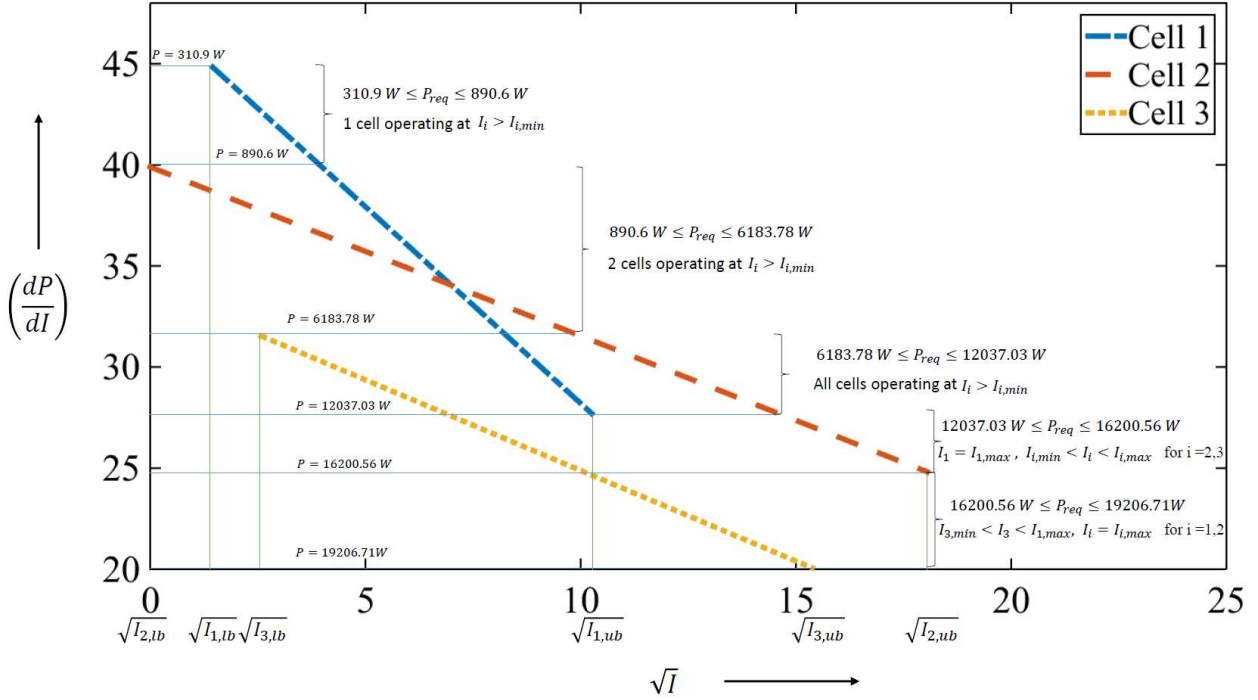}
\caption{$\nicefrac{dP_i}{dI_i}$ variation for different stacks in the network for which parameters are described in Table \ref{ex1_cellparameters}. Variation is shown with respect to $\sqrt{I}$ such that I varies from $I_{lb}$ to $I_{ub}$. Each horizontal line represents an observable point.}
\label{ex1_dpdicurve}
\end{figure*}
\subsection{Working of algorithm}
\begin{figure}[htbp]
\centering
\includegraphics[scale=0.7,trim = 0cm 0cm 0cm 0cm, clip=true]{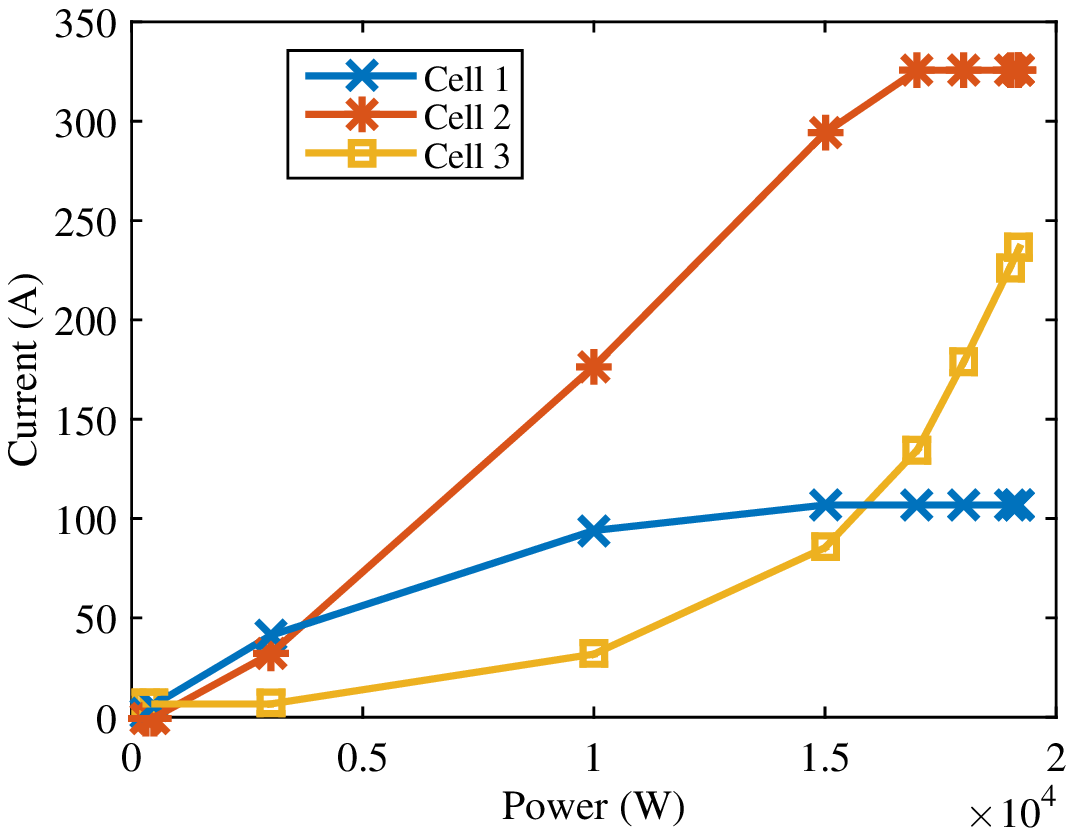}
\caption{Optimum currents obtained for stacks for various power requirement}
\label{ex1_optmcurrnt}
\end{figure}

Consider an example of a fuel cell network consisting of 3 fuel cell stacks that are connected in parallel using a power bus. Parameters used for simulation are described in Table \ref{ex1_cellparameters}. The variation of $\nicefrac{dP_i}{dI_i}$ with $\sqrt{I_i}$ for all the stacks are shown in Fig. \ref{ex1_dpdicurve}. It can be clearly seen that stack 1 has the highest $\nicefrac{dP_i}{dI_i}$ at its lower bound and so it should start operating first according to the algorithm. The observable points as explained in section \ref{algorithm} are calculated and tabulated in Table \ref{ex1_obs_ptsdata} and are also marked in Fig. \ref{ex1_dpdicurve}. 
\begin{table}[h]
  \centering
  \caption{Parameters for a fuel cell network with 3 fuel cell stacks}
    \scalebox{0.8}{\begin{tabular}{cccccc}
    \toprule
    Stack  & Lower bound & Upper bound & $\phi$  & a     & b \\
    Number & on current (A) & on current (A) & & &\\
        \midrule
    1     & 2.103 & 106.8127 & 1 & 47.655 & -1.297 \\
    2     & 0     & 325.6562 & 1 & 39.895 & -0.557 \\
    3     & 6.646 & 236.4155 & 1 & 33.847 & -0.5976 \\
    \bottomrule
    \end{tabular}}%
  \label{ex1_cellparameters}%
\end{table}%
Power obtainable from the network when all stacks are operating at their corresponding lower bounds is 310.976 W. For each observable point, stacks which are to be operated at their corresponding lower bounds and upper bounds can be pre-calculated and are given in Table \ref{ex1_obs_ptsdata}. It can be seen from both Table \ref{ex1_obs_ptsdata} and Fig. \ref{ex1_dpdicurve} that stack 1 starts operating first on increasing the power requirement from $P_{min}$ ($= 310.976 W$). When the power required reaches the power corresponding to the second observable point (corresponding to $\nicefrac{dP_2}{dI_2}$ evaluated at $I_{2,lb}$), then stack 2 also starts to operate. Each stack starts to operate one by one and for the stacks operating, optimum currents are calculated using analytical solution described in section \ref{anlyt}. If observable point corresponding to upper bound current of stack j is reached, then from that power onwards '$j^{th}$' stack should be operating at its upper bound current. This can be seen starting from observable point 4, where stacks reaches their corresponding upper bounds one by one. The process continues in a similar manner till all stacks are operating at their upper bound. Fig. \ref{ex1_dpdicurve} shows how various stacks should be operated at optima for various power requirement regimes. Fig. \ref{ex1_optmcurrnt} shows the optimum current calculated using the proposed algorithm for various power requirements.
\begin{table*}[htbp]
  \centering
  \caption{Observable points calculated for the fuel cell network with 4 stacks whose parameters are described in Table \ref{ex1_cellparameters}}
    \scalebox{0.9}{\begin{tabular}{cccccc}
    \toprule
    Sl.no & Corresponding & Corresponding  & Stacks operating& Stacks & Stacks operating\\
     & total power& dP/dI & at Imin &to be optimized & at Imax \\
    \midrule
    1     & 310.976 & 44.834 & 1,2,3 & -     & - \\
    2     & 890.577 & 39.895 & 2,3 & 1     & - \\
    3     & 6183.777 & 31.536 & 3   & 1,2   & - \\
    4     & 12037.033 & 27.548 & -     & 2,3 & 1 \\
    5     & 16200.563 & 24.818 & -     & 3 & 1,2 \\
    6     & 19206.708 & 20.064 & -     &  -  & 1,2,3 \\
    \bottomrule
    \end{tabular}}%
  \label{ex1_obs_ptsdata}%
\end{table*}%

For any power requirement, we get three solutions by solving equation \eqref{cubiceqn}. For example, when $P_{req} = 8000$, three solutions obtained are given in Table \ref{3solns}. But as explained in section \ref{anlyt}, it can be seen that only one solution satisfies upper and lower bounds for current. It is an interesting inference that only a single solution will satisfy all the constraints out of all the solutions obtained using algorithm for a convex problem. 
\begin{table}
\centering
\caption{Three solutions obtained by solving equation for the fuel cell network with parameters shown in Table \ref{ex1_cellparameters} for a power requirement of 8000 W.}
    \scalebox{0.8}{\begin{tabular}{  ccccc }
\toprule
	& Stack 1 & Stack 2 & Stack 3 & Total Current \\ \midrule
	Solution 1 & 1140.37 A & 4808.83 A & 3350.97 A & 9300.17 A\\ 
	\cellcolor{gray!25}Solution 2 & \cellcolor{gray!25}81.02 A & \cellcolor{gray!25}136.23 A & \cellcolor{gray!25}17.07 A & \cellcolor{gray!25}234.32 A\\ 
	Solution 3 & 1.93 A & 36.6 A & 153.407 A & 191.94 A\\ \bottomrule
\end{tabular}}
\label{3solns}
\end{table}
\subsection{Equivalent network formation for a network with multiple stacks in each branch}
Consider a network of 30 fuel cell stacks unequally distributed in 15 branches with parameters as given in Table \ref{unequal_scenario_1}. For applying the algorithm, an equivalent network with single stack in each branch is constructed for this network such that total power from each branch of the original network is the same as the power from that single stack in the corresponding branch of the equivalent network.

Let $P_{i}$ be the power acquired from $i^{th}$ branch in the equivalent network, then
\begin{equation}
P_i = \sum_{j=1}^{N_{f,i}} P_{ij} =\sum_{j=1}^{N_{f,i}}  \phi_{ij}V_{ij}I_i
\end{equation}
For a V-I profile given in equation \eqref{voltage profile}, 
\begin{align}
P_i &= \sum_{j=1}^{N_{f,i}}  \phi_{ij} (a_{ij} + b_{ij} \sqrt{I_i}) I_i\\
&= \sum_{j=1}^{N_{f,i}}  (\phi_{ij}a_{ij}I_i + \phi_{ij}b_{ij} I_i^{1.5})\\
&= (a_i + b_i \sqrt{I_i}) I_i
\end{align}
where $a_i = \sum_{j=1}^{N_{f,i}}  (\phi_{ij}a_{ij})$ and $b_i = \sum_{j=1}^{N_{f,i}}  (\phi_{ij}b_{ij})$ are parameters in voltage profile for the stack in $i^{th}$ branch of the equivalent network. Using the equivalent network of the above mentioned network, optimum current in each branch is estimated for a power requirement of 75000 W and is given in Table \ref{unequal_scenario_1}. 

\subsection{Comparison of proposed algorithm with conventional optimizers}
The optimum current estimated in each branch of various fuel cell networks using proposed algorithm is compared with that of standard optimizers. The optimum currents obtained using both approaches are found to be matching well for all the networks. Table \ref{unequal_scenario_1} shows the comparison of optimum current estimated using standard optimizer and proposed algorithm in Python environment for one of the networks. The time taken for optimization and the value of objective function using both approaches when implemented in Python are given in Table \ref{unequal_scenario_2}. It can be seen that the proposed algorithm is much faster than the standard optimizer. 
\begin{table*}[t]
  \centering
   \caption{Optimum current values obtained in case of unequal distribution of fuel cell stacks in different rows. $P_{req}$=75000 W,   	$\phi$=0.8, Number of Fuel Cell Stacks = 30, $0.1\leq I \leq inf$}
      \begin{tabular}{cccccc}
      \toprule
      \textbf{Stack} & \textbf{Row}   &  \textbf{a }   & \textbf{ b}   & \textbf{$I_{opt}$ (A)} & \textbf{$I_{opt}$ (A)} \\
       \textbf{No.} &   \textbf{No.}    & \textbf{values} & \textbf{values} & \textbf{{\scriptsize (Optimization routine)}}      &\textbf{{\scriptsize (Proposed Algorithm)}}  \\
     \midrule
    1     & 1     & 49.25 & -0.25 & \multirow{2}[4]{*}{9.5299} & \multirow{2}[4]{*}{9.5355} \\ 
    2     & 1     & 49.302 & -0.302 &       &  \\\midrule
    3     & 2     & 49.353 & -0.353 & 0.1   & 0.1 \\ \midrule
    4     & 3     & 49.405 & -0.405 & \multirow{3}[6]{*}{648.8999} & \multirow{3}[6]{*}{648.9024} \\
    5     & 3     & 49.457 & -0.457 &       &  \\
    6     & 3     & 49.509 & -0.509 &       &  \\ \midrule
    7     & 4     & 49.56 & -0.56 & \multirow{2}[4]{*}{3.2599} & \multirow{2}[4]{*}{3.262} \\
    8     & 4     & 49.612 & -0.612 &       &  \\ \midrule
    9     & 5     & 49.664 & -0.664 & \multirow{2}[4]{*}{2.6699} & \multirow{2}[4]{*}{2.674} \\
    10    & 5     & 49.716 & -0.716 &       &  \\ \midrule
    11    & 6     & 49.767 & -0.767 & \multirow{2}[4]{*}{2.2799} & \multirow{2}[4]{*}{2.2766} \\
    12    & 6     & 49.819 & -0.819 &       &  \\ \midrule
    13    & 7     & 49.871 & -0.871 & \multirow{2}[4]{*}{1.99} & \multirow{2}[4]{*}{1.993} \\
    14    & 7     & 49.922 & -0.922 &       &  \\ \midrule
    15    & 8     & 49.974 & -0.974 & 0.1   & 0.1 \\ \midrule
    16    & 9     & 50.026 & -1.026 & \multirow{2}[4]{*}{1.6899} & \multirow{2}[4]{*}{1.6946} \\
    17    & 9     & 50.078 & -1.078 &       &  \\ \midrule
    18    & 10    & 50.129 & -1.129 & \multirow{2}[4]{*}{1.55} & \multirow{2}[4]{*}{1.55} \\
    19    & 10    & 50.181 & -1.181 &       &  \\ \midrule
    20    & 11    & 50.233 & -1.233 & \multirow{3}[6]{*}{90.0699} & \multirow{3}[6]{*}{90.0715} \\
    21    & 11    & 50.284 & -1.284 &       &  \\
    22    & 11    & 50.336 & -1.336 &       &  \\  \midrule
    23    & 12    & 50.388 & -1.388 & \multirow{2}[4]{*}{1.3} & \multirow{2}[4]{*}{1.2976} \\
    24    & 12    & 50.44 & -1.44 &       &  \\  \midrule
    25    & 13    & 50.491 & -1.491 & \multirow{3}[6]{*}{64.1899} & \multirow{3}[6]{*}{64.1877} \\
    26    & 13    & 50.543 & -1.543 &       &  \\
    27    & 13    & 50.595 & -1.595 &       &  \\  \midrule
    28    & 14    & 50.647 & -1.647 & \multirow{2}[4]{*}{1.13999} & \multirow{2}[4]{*}{1.1365} \\
    29    & 14    & 50.698 & -1.698 &       &  \\ \midrule
    30    & 15    & 50.75 & -1.75 & 0.1   & 0.1 \\ 
    \bottomrule
    \end{tabular}%
  \label{unequal_scenario_1}%
\end{table*}%

\begin{table}[htbp]
  \centering
  \caption{Comparision of the results obtained for unequal distribution of fuel cell stacks in different rows using a standard optimization routine and the proposed algorithm. $P_{req}$=75000 W,   	$\phi$=0.8  		 Number of fuel cell stacks = 30. Parameters and optimum current drawn are described in Table \ref{unequal_scenario_1}.}
    \scalebox{0.75}{\begin{tabular}{rrcc}
    \toprule
          &       & \textbf{Optimizer} & \textbf{Proposed Algorithm} \\
    \midrule
          &       &       &  \\
    \multicolumn{2}{c}{Total Current (A)} & 828.88 & 828.88 \\
    \multicolumn{2}{c}{Total Power (W)} & 75000 & 75000 \\
    \multicolumn{1}{c}{Time taken for optimizing (sec)} & & 0.104&\textbf{ 0.0024}\\
    \bottomrule
    \end{tabular}}%
  \label{unequal_scenario_2}%
\end{table}%

Multiple trials were run for different power requirements and for various fuel cell networks where all stacks are assumed to have a concave power profile.. It was observed that in all these runs, only one solution among the three sets of solutions obtained from the algorithm satisfied upper and lower bound constraints. It is surprising to see that a cubic equation as given in \eqref{cubiceqn} will always result in one and only one solution that satisfies all the constraints for a network with all stacks having a concave power profile. For the convex subproblem, a unique optimum is guaranteed from the algorithm. 

For a general problem, if required power is feasible from the fuel cell network, then all possible local minima can be obtained from the proposed algorithm unlike conventional optimizers where a single local optimum is obtained based on the initial guess. Also, all these possible local optima are estimated in a very short time. Global optimum is guaranteed using the proposed algorithm unlike standard optimizers as it can be estimated easily using all possible local optima. If power is a concave function of current for all cells in the network, then a unique local optimum is obtained from the algorithm and this local optimum will be the global optimum to the problem. Optimum solution and computational time are highly dependent on the initial guess fed by the operator in case of any optimization routine. Whereas, the algorithm proposed does not require an initial guess and is much more computationally efficient. 

Though this problem can be solved using standard optimizers in MATLAB or Python, it is not possible to embed this optimizer in a chip due to various limitations given below. 
\begin{itemize}
	\item[(a)] It cannot determine whether a solution to the problem is feasible or not. 
	\item[(b)] It is sensitive to the initial guess passed to the optimizer.
	\item[(c)] Robustness cannot be ensured.
	\item[(d)] Increased computational load on the hardware.
	\item[(e)] It requires large computational time and storage space.
	\item[(f)] It does not guarantee global solution.
\end{itemize}
All the above limitations can be overruled by the new algorithm and thus can be embedded into a chip for on-line applications.

\section{Conclusions}
An optimization problem is formulated for the optimum power distribution control for a fuel cell network. A new and computationally efficient algorithm is proposed for this optimization problem which can easily be embedded in a chip for online applications. New algorithm is proposed based on the observations when KKT conditions are applied to the given optimization problem. The algorithm proposed here guarantees global optimum to the optimization problem concerned unlike local optimum obtained using the conventional method of using standard optimizers and in a much shorter time. This can be very useful in optimizing the current and implementing control strategies in a real fuel cell network. Analytical solution for a specific convex subproblem is also formulated in this work. Though the analytical solution is applicable only for a specific voltage current profile, the algorithm can be used for any other V-I characteristics as well, irrespective of whether the resulting optimization problem is convex or not. 

The proposed algorithm can be further extended to other power distribution systems including pumps and compressors, generators, batteries and solar cells, though a fuel cell network is considered through out this paper. For example, in a  multiple pump system used for water transport with varying demand, the algorithm can determine which pump should operate at what flow rate to satisfy the required water demand at any point of time. Here, flow rate is analogous to current while head or pressure drop is comparable to voltage. Also, electrical systems used in a parallel network configuration is similar to the fuel cell network and its optimum current distribution among connected sources, to achieve a desired power, can be identified using the algorithm. It can also be applicable for cases where combinations of different systems are connected together to form a network.  


\bibliographystyle{plain}        
\bibliography{nmrlpaper}          

\appendix
\section{Relaxed Optimization Problem} \label{red.optim.pbm}
Consider the optimization problem (P3) where the equality constraint is relaxed to inequality constraint.
\begin{subequations}
\label{optim_relaxed}
\begin{align}
& \underset{I_{i}}{\text{min}}
& & I_{net} = \sum_{i=1}^{N}I_{i}  \\
& \text{subject to} & &  \sum_{i=1}^{N} P_{i} \geq P_{req} \label{eq_const}\\
& & & I_{i,lb}-I_{i} \leq 0 \qquad \quad \forall \quad i=1:N & &\\
& & & I_{i} -I_{i,ub}\leq 0 \qquad \quad \forall \quad i=1:N & &
\end{align}
\end{subequations}
Lagrangian for this problem is 
\begin{multline}
L = \sum_{i=1}^N I_i +\lambda \left(P_{req} -\sum_{i=1}^N P_i\right) + \sum_{i=1}^{N} \mu_i (I_{i,lb} - I_i)\\ + \sum_{i=1}^{N}\gamma_i (I_i - I_{i,ub})
\end{multline}
There are a total of 2N+1 inequality constraints out of which first N are lower bounds on current, next N are upper bounds on current and the last one is the relaxed power inequalty constraint. Necessary conditions for optimality as described by KKT conditions for all $i \ \epsilon \ \{1,...,N\}$ are
\begin{gather}
\mu_i \geq 0 \\
\gamma_i \geq 0\\
\mu_i (I_{i,lb}-I_i) = 0  \\ 
\gamma_i (I_i - I_{i,ub}) = 0\\
\lambda \geq 0\\
\lambda (P_{req} - \sum_{i=1}^N P_i) = 0
\end{gather}
We get 2 additional conditions using the relaxed constraint.  Different cases using KKT conditions can be classified to mainly 2 cases as given below.
\subsection*{Case I}
Power constraint is inactive: $P_{req} < \sum_{i=1}^N P_i$ and $\lambda = 0$. 
\paragraph*{1)} For $1 \leq M \leq N$ and $1 \leq K \leq N$, \\
Constraints $1:M$ are inactive, $M+1:N$ are active.\\
Constraints $N+1:N+K$ are inactive, $N+K+1:2N$ are active.
\begin{enumerate}[(a)]
\item If $K \leq M$, for $i = M+1:N$, $I_i = I_{i,lb} = I_{i,ub}$ which is not feasible.
\item If $K > M$, for $i = K+1:N$, $I_i = I_{i,lb} = I_{i,ub}$ which is not feasible.
\end{enumerate}
\paragraph*{2)}
For $1 \leq M \leq N$ and $1 \leq K \leq N$, \\
Constraints $1:M$ are inactive, $M+1:N$ are active.\\
Constraints $N+1:N+K$ are active, $N+K+1:2N$ are inactive.
\begin{enumerate}[(a)]
\item If $K>M$, 
For $i = 1+M:K$, $I_i = I_{i,lb} = I_{i,ub}$ which is not feasible.
\item If $K\leq M$,
\begin{equation*}
I_i = \left\{ \,
\begin{tabular}{c c}
$I_{i,ub} $ & for $i=1:K$ \\
$I_{i,lb}$ & for $i=1+M:N$ 
\end{tabular}
\right.
\end{equation*}
Using $\frac{dL}{dI_i} = 0$, 
\begin{equation}
\label{dldi_inequal}
1 - \mu_i +\gamma_i = 0 \qquad \forall \ i=1:N
\end{equation}
For the given case, KKT conditions can be reduced to
\begin{gather*}
\mu_i = \left\{ \,
\begin{tabular}{c c}
0 & for $i=1:M$ \\
$\geq 0$ & for $i=1+M:N$ 
\end{tabular}
\right.\\
\gamma_i = \left\{ \,
\begin{tabular}{c c}
$\geq 0$ & for $i=1:K$ \\
0 & for $i=1+K:N$ 
\end{tabular}
\right.
\end{gather*}
\end{enumerate}
Substituting in equation \eqref{dldi_inequal}, 
\begin{align}
1 + \gamma_i &= 0 \qquad \forall \quad i = 1:K \\
1 &= 0 \qquad \forall \quad i = 1+K:M \\
1 - \mu_i &= 0 \qquad \forall \quad i = 1+M:N 
\end{align}
which cannot be satisfied. Hence, this case is also not a feasible solution.
\subsection*{Case II}
Power constraint is active: $P_{req} = \sum_{i=1}^N P_i$ and $\lambda \geq 0$. 
The sub cases when power constraint is active is same as the cases given in section \ref{genrlsoln} and we get the same conditions for feasible solution as the original optimization problem (P2) given by 
\begin{multline}
\left( \frac{dP_{k+1}}{dI_{k+1}}\right)_{I_{k+1,opt}}=\left( \frac{dP_{k+2}}{dI_{k+2}}\right)_{I_{k+2,opt}}=... \\ =\left( \frac{dP_{M}}{dI_{M}}\right)_{I_{M,opt}}  
\end{multline}
and 
\begin{gather}
\left( \frac{dP_j}{dI_j} \right)_{I_{j,lb}} \leq \left( \frac{dP_i}{dI_i} \right)_{I_{i,opt}} \leq \left( \frac{dP_r}{dI_r} \right)_{I_{r,ub}} \\
j = M+1:N; \ \ i = K+1:M; \quad r = 1:K
\end{gather}
Hence, the relaxed optimization problem has a feasible solution if and only if the power constraint is active. Moreover, the solution set is the same for both minimization problems P2 and P3. Thus, those two optimization problems can be considered equivalent.

\subsection{Convex Subproblem}
Let the power of each stack ($P_i = \phi_i \ I_i \ f(I_i)$) be a concave function (or in other words Hessian of power be negative definite), then all the constraints in optimization problem (P3) given by equation \eqref{optim_relaxed} together form a convex set. Moreover, the objective function is linear. Thus, the minimization problem (P3) is a convex problem for a concave power function.

The solution to this convex problem can be obtained from the same conditions derived in section \ref{genrlsoln}. But, as it is a convex problem, only one minima is possible while equations \ref{equaldpdi} and \ref{allcondnskkt} will result in multiple solutions. Hence, among all possible multiple solutions to those conditions, only one of the solutions will satisfy the bounds on current and results in a single optima to the convex subproblem. 
\end{document}